\documentclass[a4paper,12pt]{article}

\usepackage{graphicx}
\usepackage{amssymb}

\begin{document}

\title{Acceleration of low energy charged particles by gravitational waves}
\author{G. Voyatzis, L. Vlahos, S. Ichtiaroglou, D. Papadopoulos}

\maketitle
University of Thessaloniki, Dept of Physics, 54124 Thessaloniki, Greece, email: voyatzis@auth.gr

\begin{abstract}
The acceleration of charged particles in the presence of a magnetic field and gravitational waves is under consideration. It is shown that the weak gravitational waves can cause the acceleration of low energy particles under appropriate conditions. Such conditions may be satisfied close to the source of the gravitational waves if the magnetized plasma is in a turbulent state.
\end{abstract}

\vspace{0.5cm}
{\bf to appear in } {\em Phys.Lett.A}

\section{Introduction}

The interaction of a charged particle with a gravitational wave (GW) has been studied  by using a Hamiltonian formalism (\cite{VP}-\cite{KVP3}). This approach is similar to the one used for the study of the interaction of ions with electrostatic waves in magnetized plasmas (\cite{Fuku}-\cite{SmK}). The existence of chaos in phase space has been shown as a consequence of the overlap of resonances.
Recently, the acceleration of the charged particles due to the chaotic diffusion has been associated with bursts \cite{VVP}. However, such a diffusion is obtained for relatively high energy particles and becomes efficiently present only for large amplitude values of the GW ($a>0.1$).

Furthermore, the existence of a parametric resonance in the general problem of interaction between particles and waves  is discussed in the classical book of Landau and Lifshitz  \cite{LaLi}. For the particular problem Kleidis {\it et al.} \cite{KVP3} showed that a parametric resonance arises for a set of frequencies of the GW  of the form $\nu=2/n,\: n=1,2,...$ and an estimation of its width is given. Also, the parametric resonance has been associated with the existence of chaotic motion.

In the present paper we study the possible fast acceleration of low energy particles for small amplitudes of the GW by taking into account two special properties of the system. The first is the parametric resonance which, as we shall show, arises only for $\nu=2$. The second property is the integrability of the system, and subsequently the absence of chaos, when the GW propagates in a parallel direction with that of the magnetic field.

In section 2 we review the Hamiltonian formalism of the model which describes the interaction. In section 3, we study systematically the rise of a parametric resonance and in section 4 we prove the integrability of the system in the case of parallel propagation. It is shown in section 5 that when the above integrable case is in parametric resonance fast acceleration of particles occurs. The astrophysical implications are discussed in section 6.

\section{Review of the system. The Hamiltonian formalism}

The dynamics of a charged particle in a curved space-time and in the presence of a uniform magnetic field can be studied by considering the Hamiltonian (see \cite{Misner})
\begin{equation}
H(x^k,p_k)=\frac{1}{2}g^{ij}(p_i-eA_i)(p_j-eA_j)=\frac{1}{2},\;\;
i,j,k=0,...,3, \label{Hoo}
\end{equation}
where $x^k$  and $p_k$ denote the generalized coordinates and momenta of the particle, respectively. The metric $g^{ij}$ defines the curved space-time and $A_i=A_i (x^k)$ are the components of the vector potential $\vec{A}$ of the magnetic field. It can be considered that the curvature of space-time is due to a gravitational wave (GW). By considering a linearly polarized GW propagating in a direction $\vec{k}$ of angle $\theta$ with respect to the axis $z$, a possible form of the tensor $g^{ij}$ is the following \cite{Ohanian,PapEsp}
\begin{equation}
g^{ij}=\eta^{ij}+a h^{ij},\:\:\:a\ll0,
\end{equation}
where $\eta^{ij}=\{1,-1,-1,-1\}$ is the Minkovski space-time and $a$ is the amplitude of the gravitational wave. By using the metric tensor derived in \cite{PapEsp} and expanding its components up to $O(a)$ we obtain the nonzero components of the tensor $h^{ij}$ in the following form 
\begin{equation}
h^{11}=-\Psi\cos^2\theta,\quad h^{22}=\Psi,\quad h^{33}=-\Psi\sin^2\theta,\quad h^{13}=h^{31}=\frac{1}{2}\Psi\sin(2\theta),
\end{equation}
where $$\Psi=\cos\left ( \nu (x^1\sin\theta+x^3\cos\theta-x^0) \right ).$$ The constant parameter $\nu$ is associated with the frequency $\omega$ of the GW. In the case of a uniform magnetic field $\vec{B}=B_0 \vec{e}_z$, which corresponds to $\vec{A}=\{0,0,B_0 x^1,0\}$, we may choose a system of units where the Larmor frequency of the particles is $\Omega=e B_0/mc=1$. In this case $\nu$ is the ratio of the GW frequency to the Larmor frequency i.e. $\nu=\omega/\Omega$.

By substituting in (\ref{Hoo}) the above specific expressions and performing a canonical transformation to  new canonical variables ($\phi,q,J,p$) (for details see \cite{VVP}) we get 
\begin{equation}
H=\frac{1}{2} \left ( I^2-2(I+p\sin\theta)J-(p^2+q^2) \right ) + \frac{1}{2} a (q^2- p^2\cos^2\theta) \cos(\nu \phi) =\frac{1}{2}
\label{Ham1}
\end{equation}
The variables $p,q$ are associated with the gyro motion of the particles in the magnetic field, $J=1-\gamma$, where $\gamma=(1-\upsilon^2)^{-1/2}$ is the particle's energy and $\upsilon$ its velocity. The quantity $I=p_0+p_3/\cos\theta$ is a constant of motion due to the symmetries of the original system.

\section{The basic periodic solution, stability and parametric resonance}
For the Hamiltonian (\ref{Ham1}) we can easily obtain the periodic solution
\begin{equation}
p(t)=q(t)=J(t)=0, \qquad \phi(t)=-I t + \phi_0\;\; (\textrm{mod} 2\pi)
\label{PO}
\end{equation}
The solution (\ref{PO}) corresponds to rest particles since it provides $\gamma=1$ (the energy of the rest mass). Because of the restriction $H=1/2$ we should set $I=1$ in order our system to possess the above special solution.
Thus, the Hamiltonian is now written in the form
\begin{equation}
H=J+ p J \sin\theta+\frac{1}{2}(p^2+q^2)- \frac{1}{2} a (q^2-p^2\cos^2\theta) \cos(\nu \phi) =0
\label{Ham2}
\end{equation}
and the solution (\ref{PO}) is periodic of period $T=2\pi$.

By considering the variations $\xi_1,\xi_2,\xi_3,\xi_4$ corresponding to the variables $\phi,q,J$ and $p$, respectively, the variational equations for the periodic solution (\ref{PO}) are 
\begin{eqnarray}
\dot{\xi_1}=\sin\theta\; \xi_4, & \qquad \dot{\xi_2}=\sin\theta \;
\xi_3+(1+a \cos^2\theta \cos(\nu t)) \xi_4, \\ \nonumber
\dot{\xi_3}=0, & \qquad \dot{\xi_4}=(-1+a \cos(\nu t)) \xi_2
\label{VEQS0}
\end{eqnarray}

Furthermore, let $\mathbf{M}$ be the monodromy matrix. Considering the unperturbed system (i.e. $a=0$) we obtain from the monodromy matrix $\mathbf{M}$ the stability index
\begin{equation}
k=2-\textrm{trace}(\mathbf{M})=2-4\cos \left ( \frac{\pi}{\nu} \right )^2
\label{Stab1}
\end{equation}
The condition $-2<k<2$, ensures the linear stability of the periodic orbits for all values $\nu$ except for $\nu=2/n,
n=1,2,...$. These values are candidates for the generation of a parametric resonance.

In the perturbed system we examine the stability by following the methodology described by Yakubovich and Starzhinskii \cite{YaSta}, p.357. The variational equations (\ref{VEQS0}) are of the form 
\begin{equation}
\dot{\vec \xi}=(\mathbf{K}_0+a \mathbf{D}_1(t)) \vec\xi
\label{VEQS1}
\end{equation}
and the fundamental $4\times 4$ matrix of solutions is
\begin{equation}
\mathbf{\Xi}(t;a)=\mathbf{F}(t;a)\, e^{\mathbf{K}(a)\,t}\;,\;\;\; \mathbf{K}(a)=\mathbf{K}_0+a \mathbf{K}_1+a^2 \mathbf{K}_2+...
\label{VEQSSOL}
\end{equation}
where $\mathbf{F}(t;a)$ is a $T$-periodic matrix, $\mathbf{K}_0$ is the matrix of coefficients of the linear system (\ref{VEQS0}) and $\mathbf{K}_i,\; i=1,2,...$ are constant matrices calculated from the variational equations (\ref{VEQS1}) on the periodic orbit. After some calculations we find $\mathbf{K}_i=0, \forall i\neq 0,2$. The nonzero components of the matrix $\mathbf{K}_2=\{K_2(i,j)\}$ are
$$
K_2(2,3)=\frac{1}{2} \frac{c (3 b^2 +\nu^2-1)}{\nu^4-5\nu^2+4},\qquad
K_2(2,4)=\frac{1}{2}\frac{b(1+b)}{\nu^2-4},\qquad
K_2(4,2)=-\frac{1}{2}\frac{(1+b)}{\nu^2-4},
$$
where $b=\cos^2\theta$ and $c=b \sin\theta$. The stability of the periodic solution is given by the eigenvalues of the matrix $\mathbf{K}(a)$ defined in (\ref{VEQS1}), which are
\begin{equation}
\lambda_1=\lambda_2=0,\;\; \lambda_3=-\lambda_4=i \left ( 1-\frac{a^2 (1+b)^2}{4(\nu^2-4)} \right )+O(a^4)
\label{EigVal}
\end{equation}
Conclusively for every value of $a$,$\theta$ and $\nu\neq 2$ the eigenvalues $\lambda_{3,4}$ are purely imaginary and  the periodic orbit that corresponds to the rest particles is stable. Therefore, for arbitrarily small amplitude values $a$, instability (parametric resonance) may occur only at $\nu=2$. At this value the above method fails to classify the periodic orbit and we proceed to a numerical investigation. Namely, we obtain the monodromy matrix by integrating numerically the variational equations beside the canonical equations of the Hamiltonian (\ref{Ham2}) for various values of the amplitude $a$ of the gravitational wave and of the propagation angle $\theta$.
\begin{figure}[t]
\centering
\includegraphics[width=8cm]{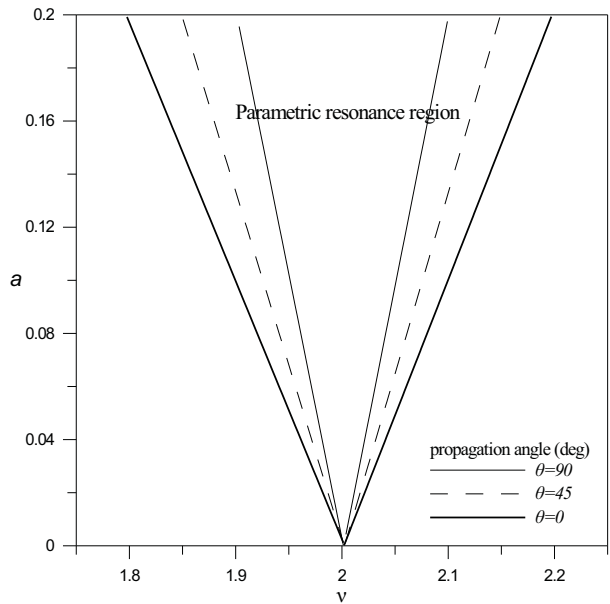}
\caption{The region of the parametric resonance in the parameter space $\nu-a$ for some typical values of the propagation angle $\theta$.}
\label{FF1}
\end{figure}
\begin{figure}[b]
\centering
\includegraphics[width=13cm]{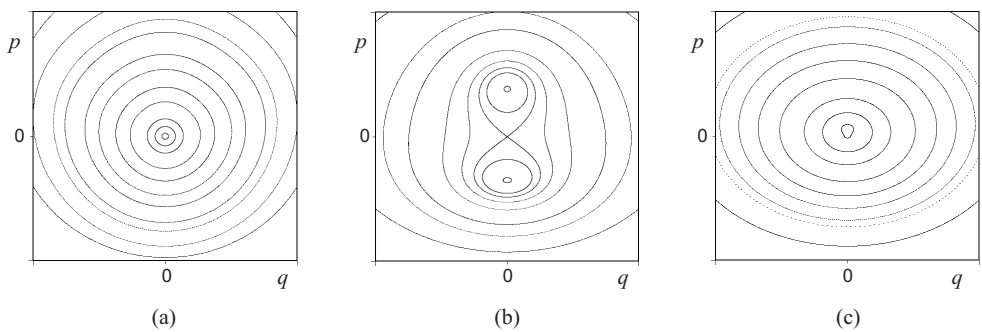}
\caption{Poincare sections of the phase space around the periodic orbit ($q=p=0$) for $a=0.01$, $\theta=30^\circ$ and (a) $\nu=4$,  (b) $\nu=2$, (c) $\nu=1$.}
\label{FF2}
\end{figure}

As it is shown in figure 1, the periodic orbit becomes unstable in a V-shaped region of parameters centered at $\nu=2$. We observe that the borders of the parametric resonance can be approximated by the fitting formula
$$a= (1+\sin^2 \theta) \Delta \nu, \:\:\Delta \nu =|\nu-2|$$
and the instability region broadens as $\theta \rightarrow 0$, i.e. towards the parallel propagation. In figure 2 we present the Poincare sections of (\ref{Ham2}) on the plane $q-p$ for $\phi=0\quad (\textrm{mod} (2\pi/\nu))$. The periodic orbit correspond to the fixed point of the sections.  We see that at the parametric resonance at $\nu=2$ a pitchfork bifurcation of the periodic orbit (\ref{PO}) occurs. Two stable periodic orbits bifurcate, which correspond to $\gamma\neq 0$, while the periodic orbit (\ref{PO}) becomes unstable and the associated stable and unstable manifolds form a separatrix curve of figure-eight.

The Poincare section of system (\ref{Ham2}) can be defined also in the plane of variables $\phi - \gamma$ for $q=0, H=0$. Then the topology of the section in Fig.2b is represented in the section $\phi - \gamma$ as it is shown in Fig.3a. The unstable periodic orbit (\ref{PO}), indicated by P$_0$, appears at $\phi=\pm \pi/2$ and its separatrix curve is located in the enegry domain $0\leq \gamma <\gamma_S$. For any initial conditions the particle's energy $\gamma$ shows regular bounded oscillation of librational type around $P_1$ or rotational type above the separatrix $S$. The stable periodic orbit, which is generated by the pitchfork bifurcation and indicated as P$_1$, corresponds to energy value $\gamma_1$. As it is illustrated in Fig.3b, the value $\gamma_1$ depends on both parameters $a$ and $\theta$. It holds
always $\gamma_1<\gamma_S$ and also we can obtain by fitting that $\gamma_1 \sim \theta^{-2}$ for $\theta\rightarrow 0$.
Consequently, as $\theta \rightarrow 0$, the energy oscillations increase rapidly their amplitude and at $\theta=0^\circ$ all particles are ejected to infinite energy independently of the amplitude of the GW. This special case is studied in the following section. 

\begin{figure}[tb]
\centering
\includegraphics[width=1\textwidth]{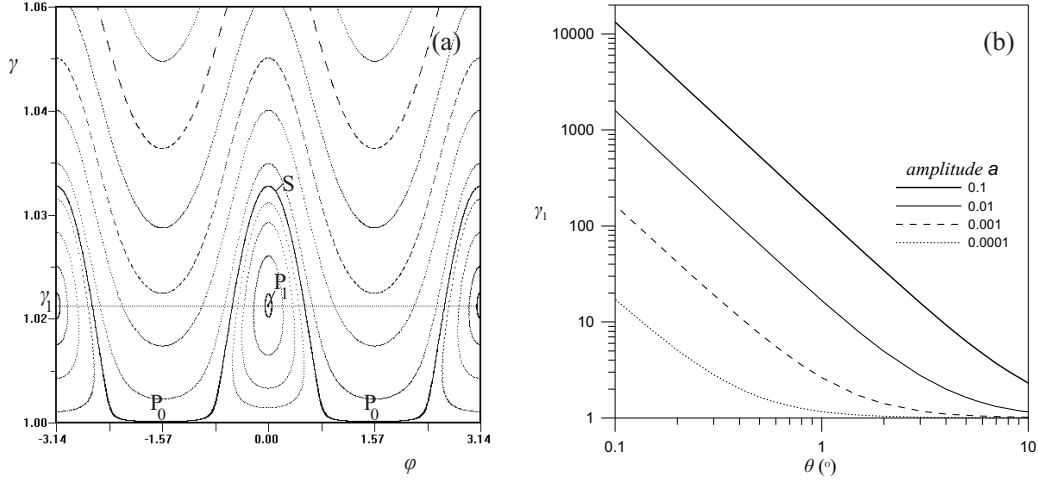}
\caption{(a) Poincare section presented in the $\phi-\gamma$ plane for $\nu=2$, $a=0.01$ and $\theta=30^\circ$. (b) The energy $\gamma_1$ of the bifurcated stable periodic solution as a function of the propagation angle $\theta$. As $\theta \rightarrow 0$ we obtain that $\gamma_1 \sim \theta^{-2}$.}
\label{FF3}
\end{figure}

\section{The case of parallel propagation}
We consider $\theta=0^\circ$, i.e. the GW propagates in parallel direction with the magnetic field. Then the Hamiltonian (\ref{Ham2}) takes the form
\begin{equation}
H=J+\frac{1}{2}(p^2+q^2)+\frac{a}{2}(p^2-q^2)\cos (\nu \phi)=0
\label{Ham20}
\end{equation}
The corresponding canonical equations are:
\begin{eqnarray}
\dot{\phi}&=&1 \label{H0a}\\
\dot{J}&=&\frac{a \nu}{2}(p^2-q^2)\sin (\nu \phi) \label{H0b}\\
\dot{q}&=&p(1+a \cos {\nu \phi}) \label{H0c} \\
\dot{p}&=&q(-1+a \cos {\nu \phi}) \label{H0d}
\end{eqnarray}

From equation (\ref{H0a}) we obtain directly that $\phi=t+\phi_0$ and subsequently equations (\ref{H0c}) and (\ref{H0d}) consist a linear non-autonomous system. Providing a solution $p=p(t), q=q(t)$ the system turns to be integrable for any values of the parameters $a$ and $\nu$. The Poincare section in Fig.4a, indicates the regular oscillations of the particle's energy for all initial conditions. Such a situation changes at the parametric resonance for $\nu=2$ and the dynamics is depicted by the Poincare section in Fig.4b. We obtain that the stable and the unstable manifolds which originate from the unstable periodic orbit (\ref{PO}) are located at $\phi=(2k-1)\pi/4$ and $\phi=(2k+1)\pi/4$ ($k\in Z$), respectively, and extend from $\gamma=1$ up to infinity. The dynamics becomes purely hyperbolic.

\begin{figure}[t]
\centering
\includegraphics[width=13.5cm]{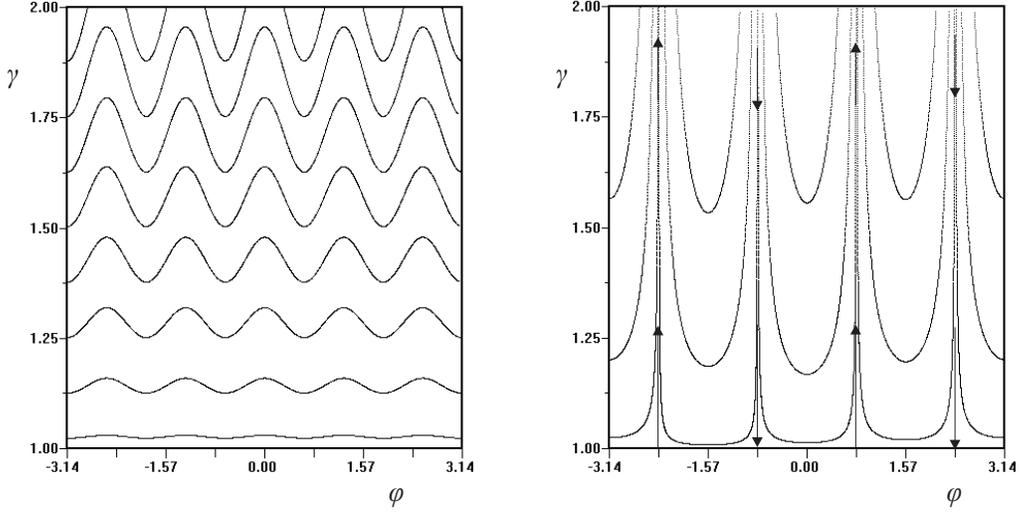}
\caption{Poincare sections in the case of parallel propagation ($\theta=0^\circ$) for (a) $\nu=5$, $a=0.1$ and (b) $\nu=2$, $a=0.001$. The up and down directed arrows indicate the unstable and stable manifolds, respectively.}
\label{FF4}
\end{figure}

The linear system of eqs. (\ref{H0c}) and (\ref{H0d}) is not directly solvable. By differentiating and introducing the variable transformation $u=p+q$ and $v=p-q$, we derive the separable system of second order equations
\begin{eqnarray}
\ddot{u}+(1+a \nu \sin (\nu t) ) u =a^2 \cos^2(\nu t) u  \label{Udeq} \\
\ddot{v}+(1-a \nu \sin (\nu t) ) v =a^2 \cos^2(\nu t) v.  \label{Vdeq}
\end{eqnarray}
If we introduce the new time variable $\tau=\pi/4-\nu t/2$ and omit the terms $O(a^2)$ the above equations take the form
\begin{equation}
\chi''+\left (\frac{4}{\nu^2} \pm \frac{4 a}{\nu} \cos (2 \tau) \right ) \chi=0,
\label{Meq}
\end{equation}
where $\chi$ stands for either $u$ or $v$, the minus sign refers to $v$ and the plus sign to $u$ and prime denotes differentiation with respect to $\tau$. Equation (\ref{Meq}) is a Mathieu type equation with solutions of the form $\chi(\tau)=e^{\lambda \tau} w(\tau)$, where $w(\tau)$ is a periodic function \cite{Abr}. In the parameter space $\nu-a$ stable and unstable regions are defined where $Re[\lambda]=0$ and $Re[\lambda]\neq 0$ respectively. For the actual system (\ref{Ham20}) the unstable region is the V-shaped region shown in Fig. 1 for $\theta=0^\circ$. In Fig.5a we present an example of the oscillations of the variable $q$ which correspond to the trajectory along the unstable manifold
for $\nu=2$. Such a variation is associated with an increase of the particle's energy $\gamma=1-J$. From (\ref{Ham20}) it follows that 
\begin{equation}
\gamma=1+\frac{1}{2}(p^2+q^2) + O(a)=1+\frac{1}{4}(u^2+v^2) + O(a)
\label{gamma0}
\end{equation}
The increase of $\gamma$ is exponential (see Fig. 4b) and the numerical study suggest that for $t\gg 1$, the associated exponent coincides with the amplitude of the GW, i.e.
\begin{equation}
\gamma \sim e^{a t}\qquad \textrm{for}\qquad   t\gg 1
\label{gammaExp}
\end{equation}

\begin{figure}[t]
\centering
\includegraphics[width=12cm]{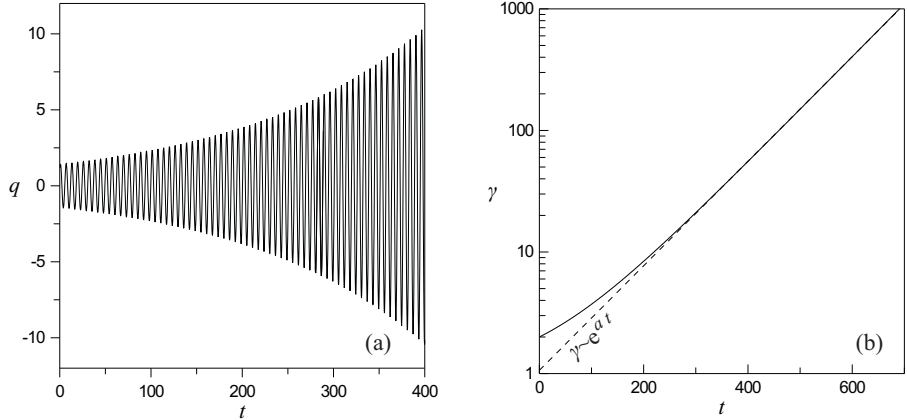}
\caption{(a) The increasing amplitude $q=q(t)$ of the gyro motion of a particle in the parametric resonance (b) the exponential increase of energy $\gamma=\gamma(t)$ for the evolution along the unstable manifold. The corresponding parameters are $\nu=2$,  $a=0.01$ and $\gamma(0)=2$.}
\label{FF5}
\end{figure}

\section{Passage through the parametric resonance}
In previous studies, mentioned in the introduction, the acceleration of the particles is a result of the
chaotic diffusion. However, wide chaotic regions are formed only for high energy levels and in the presence of very strong gravitational waves ($a>0.01$) \cite{VVP}. Especially, for small values of the propagation angle $\theta\neq 0$ chaotic regions appear in phase space only as very thin bounded zones around islands formed according to the Poncare-Birkhoff theorem. Consequently, acceleration of low energy particles under the perturbation caused by weak GWs can occur only through the parametric resonance ($\nu=2$) at the integrable case ($\theta=0^\circ$).

In a more realistic situation, the direction and the magnitude of the magnetic field met by the particles is not constant. According to the results obtained in the previous sections, the charged particles will accelerate when the magnetic field becomes almost parallel to the direction of the gravitational wave and have a magnitude such that $\nu\approx 2$. A simple simulation that illustrates this situation is described below.

We consider that the propagation angle between the GW, with constant $\nu=2$, and the direction of the magnetic field varies in time linearly as 
\begin{equation}
\theta(t)=\theta_0+\lambda t
\label{VTH}
\end{equation}
The time variable $t$ is the coordinate time $x^0$ in the original form (\ref{Hoo}) of the system. The slope parameter $\lambda$ controls the time span spent by the particle in an interval ($\theta_1,\theta_2$) and, particularly, the critical time interval $\Delta t^*$ where $\theta\approx 0$. A typical example of the evolution of a particle under the above condition is shown in Fig.6a.

\begin{figure}[t]
\centering
\includegraphics[width=1 \textwidth]{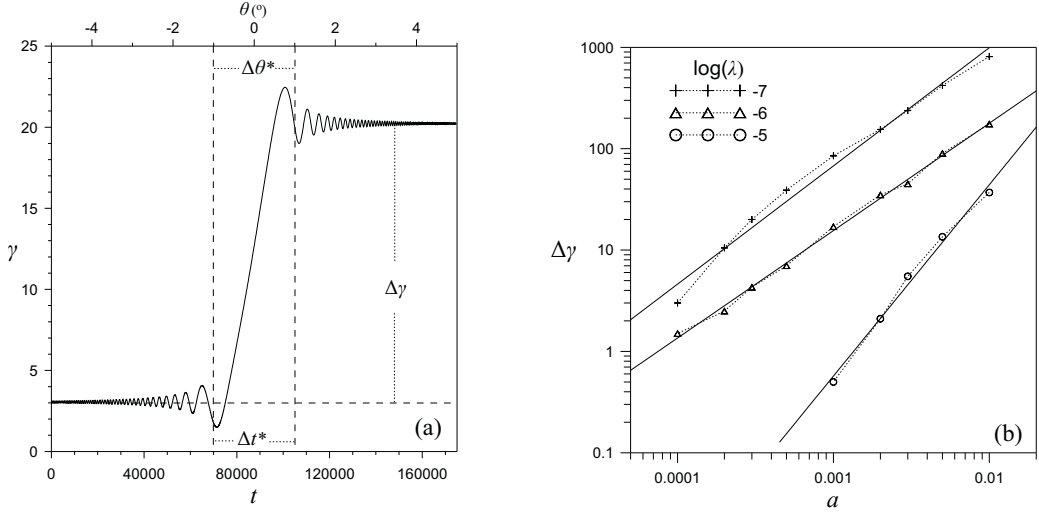}
\caption{(a) The evolution of $\gamma$ along a trajectory in the system with varying propagation angle, $\nu=2$ and $a=0.001$. The energy jump is clearly seen when $\theta\approx 0^\circ$. (b) The magnitude of the energy jump $\Delta \gamma$ against the amplitude of the GW.}
\label{FF6}
\end{figure}

We observe that the particle, which starts with low energy $\gamma=\gamma_0$ (particularly $\gamma_0=3$) and 		$\theta_0=-5^\circ$, obeys small oscillations for approximately $\theta<-1^\circ$. When the system approaches the integrable hyperbolic dynamics at $\theta=0^\circ$, the particle accelerates and a significant jump in particle's energy occurs. Leaving the hyperbolic domain ($\theta>1$) the particle is captured in a regular phase space domain with corresponding energy $\gamma=\gamma_1$ significantly greater than $\gamma_0$ (particularly, $\gamma_1\approx 20$). The total increase in energy, $\Delta \gamma$, depends on the critical time interval $\Delta t^*$ and the $\theta$-width, $\Delta \theta^*$, where the hyperbolic structure is effective. For the particular simple model it holds that $\Delta t^*=\Delta \theta^*/\lambda$. Certainly, the most significant parameter for the acceleration is the amplitude $a$ of the GW which is associated with the exponential rate of the acceleration along the unstable manifolds in phase space according to the estimation (\ref{gammaExp}). In Fig.6b we present the relation between $a$ and $\Delta \gamma$ derived by the numerical simulation for some particular values of the parameter $\lambda$. We observe that, if the critical time $\Delta t^*$ is sufficiently large (or $\lambda$ is small),  the energy jump is notable even for very small values of $a$. The same qualitative results are obtained when we keep constant the propagation angle  $\theta=0^\circ$ and change the parameter $\nu$ passing through the value $\nu=2$.  

\section{Astrophysical implications and conclusions}

We have shown that gravitational waves with frequency $\omega$ and amplitude $\alpha$ can accelerate charged
particles to very high energies when two conditions are met (a) the gravitational wave propagates along the external magnetic field ($\theta\approx 0^\circ$) and (b) the ratio of the gravitational wave frequency to gyro frequency is about 2 ($\nu\approx 2)$. The model used has  a number of important simplifications: (1) We assume that the magnetic field is constant (2) The energy absorption by the particles is small compared to the energy curried by the gravitational wave and (3) we ignore the collective plasma effects associated with the particle acceleration.    

Retaining the above simplifications we may proceed to estimate the energy gained by the particles in a realistic astrophysical system. We assume that a volume of length $L \approx 10^{12} cm$ is close to a source of gravitational waves. The volume is filled with plasma with density $n\approx 10^{12} \textrm{particles}/cm^{-3}$ and the
magnetic field is in a fully turbulent state. The gravitational wave is crossing  the volume in few seconds and
every time it travels along the ambient field ($\theta \approx 0^\circ$) with almost zero strength its energy increases. The zero strength magnetic field, which is associated to null surfaces, is an important requirement since the gravitational wave has low frequency (of few KHz) and acceleration is possible only for $\nu\approx 2$).  The energy increase is fast and can be substantial ($\Delta\gamma\approx 10$) if the amplitude of the gravitational wave is $a\approx 10^{-4}-10^{-5}$. Therefore the particles diffuse in energy space not by small random steps, as it was the case in the stochastic interaction studied earlier (\cite{VP}- \cite{KVP3}), but through large random energy jumps. 

It is beyond the scope of this article to present a detailed analysis of the above process but we can present a rough estimate of the energy transferred from the gravitational wave to the plasma. Let us assume that a particle travel a distance $L_p\geq L$ (its trajectory is not a straight line) before exiting the interaction volume. Let  us also assume that only in small parts of its trajectory ($L_{int}\approx $ a few kilometers) the particle is able to meet the conditions needed for its acceleration. For a small portion of protons, e.g $n_{acc}=n_{total}/100$), which participate in the interaction, these regions can be a fraction of $N_{max}=L_p/L_{int}\approx 10^7$ and the energy gain will be 
$E_{kin} \approx 10^6 (\Delta \gamma-1) m_p c^2 \approx 10^4 ergs$ (or $10^{16} eV$), where $m_p$ is the mass of proton. 
The number of particles participating in this interaction can be as many as $N_{total} \approx n_{acc} L^3 \approx 10^{46}.$ The total energy lost by the gravitational wave is, approximately, $10^{48}-10^{49}$ ergs, which
constitute a small fraction of its total energy $10^{58}$ ergs (see \cite{Putten}).

According to the model, the acceleration conditions ($\theta\approx 0^\circ$, $\nu\approx 2$) can be met sporadically along the trajectory of a charged particle in the vicinity of a gravitational wave source when the magnetized plasma is in a turbulent state. A few particles can then reach very high energies in a few seconds
absorbing only a very small fraction of the gravitational wave. It remains to be seen, by performing further simulations, if the process described above for particle acceleration can be significant in a realistic astrophysical environment. In this case the electromagnetic emission from the accelerated particles can be a precursor of the gravitational wave for easier detection.

\end{document}